\documentclass[prb,twocolumn,floatfix]{revtex4}
\usepackage{amssymb}
\usepackage{amsmath}
\usepackage{graphicx}

\begin{document}

\title{Effective charging energy of the single electron box}
\author{Philipp Werner}
\affiliation{Institut f{\"u}r theoretische Physik, ETH H{\"o}nggerberg, CH-8093 Z{\"u}rich, Switzerland}
\author{Matthias Troyer}
\affiliation{Institut f{\"u}r theoretische Physik, ETH H{\"o}nggerberg, CH-8093 Z{\"u}rich, Switzerland}
\date{\today}

\begin{abstract}
We present numerical results on electron tunneling in a single-electron box at low temperature.
The effective action of this device is equivalent to the Hamiltonian of a classical XY spin chain with long ranged interactions. Using an efficient cluster algorithm and a new transition matrix Monte Carlo approach, we are able to compute the effective charging energy $E_C^*$ in the limit of very small tunneling resistance. While previous Monte Carlo simulations were restricted to the weak and intermediate tunneling regimes, our method extends the range of  $E_C^*$-values by more than 30 orders of magnitude. This allows us to clearly observe the exponential suppression of $E_C^*$ with increasing tunneling conductance $\alpha$. For large, but fixed $\alpha$, the correction to the leading exponential behavior exhibits a crossover from an intermediate temperature behavior at $\beta E_C^* \ll 1$, to zero temperature behavior at $\beta E_C^*\gg 1$. We determine this correction in both regimes and compare the numerical results to the numerous and controversial theoretical predictions for the strong tunneling limit.
\end{abstract}

\maketitle

\section{Introduction}

The so-called single electron box is an example of a dissipation coupled mesoscopic system with discrete charge states. It consists of a low-capacitance metallic island connected to an outside lead by a tunnel junction. The single electron box has been the subject of numerous experimental \cite{Lafarge} and theoretical \cite{Goeppert, Panyukov_Zaikin, Wang_Grabert, Lukyanov, Hofstetter, Koenig} investigations, because it exhibits Coulomb blockade phenomena due to the large charging energy of the island. The presence of excess charges influences single electron tunneling and this effect is used in single electron transistors\cite{Averin, Fulton}, which can be viewed as a single electron box connected to two tunnel junctions for the entrance and exit of electrons.

\begin{figure}[h]
\centering
\includegraphics [angle=0, width= 5.2cm] {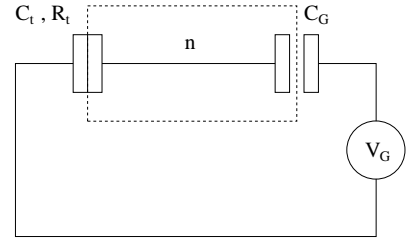}
\caption{Circuit diagram of the single electron box. The box with excess charge $n$ is indicated by the dashed rectangle. It is connected to a voltage source through a capacitor $C_G$ and a tunnel junction with resistance $R_t$ and capacitance $C_t$.}.
\label{box}
\end{figure}

The first experimental realization of a single electron box was achieved by Lafarge \textit{et al.} \cite{Lafarge}. A circuit diagram of this device is shown in Fig.~\ref{box}. The box with excess charge $n$ is controlled by an external voltage source $V_G$ to which it is connected through a capacitor $C_G$ and a tunnel junction with resistance $R_t$ and capacitance $C_t$. An applied gate voltage $V_G$ induces a continuous polarization charge $n_G=C_GV_G/e$. The bare charging energy $E_C=e^2/2(C_t+C_G)$ sets the energy scale. 

\begin{figure}[h]
\centering
\includegraphics [angle=0, width= 7.5cm] {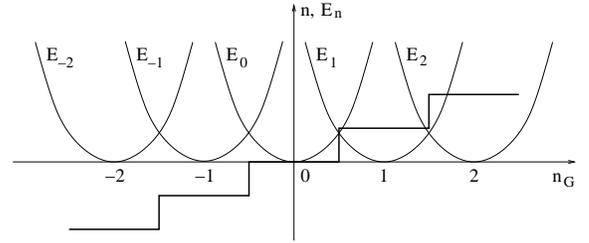}
\caption{Illustration of the Coulomb staircase in the limit $k_BT \ll E_C$ and $R_t \gg R_K$.}.
\label{step}
\end{figure}

The classical electrostatic energy for an integer number $n$ of additional electrons on the island is given by the set of parabolas $E_n(n_G) = E_C(n-n_G)^2$, as shown in Fig.~\ref{step}. In the ground state, the number of excess charges on the island is therefore a staircase function with unit jumps at $n_G=1/2$ (mod 1).    

This simple classical picture is valid at low temperatures, $k_BT \ll E_C$, and high tunneling resistance, $R_t \gg R_K=h/e^2\approx 25.8 k\Omega$. Thermal fluctuations will round off the corners of the step structure. But even at zero temperature, electron tunneling processes can smear out the staircase, which for large tunneling approaches a straight line. These quantum fluctuations renormalize the ground state energy and lead to an effective charging energy

\begin{equation}
E^*_C=\left.\frac{1}{2}\frac{d^2}{d{n_G}^2}E_0(n_G)\right|_{n_G=0}.
\end{equation}  

Several recent theoretical papers have addressed the renormalization of the charging energy due to electron tunneling processes. Perturbative expansions, valid for small values of the dimensionless tunneling conductance $\alpha=\frac{1}{2\pi^2}\frac{R_K}{R_t}$, yield \cite{Goeppert}

\begin{equation}
\frac{E_C^*}{E_C}=1-2\alpha+1.27\alpha^2-0.182\alpha^3+O(\alpha^4).
\label{perturbation_series}
\end{equation} 
In order to treat the opposite limit of large tunneling conductance, various approaches have been proposed -- with controversial results. While these analytical predictions agree on the leading exponential suppression of the effective charging energy, 

\begin{equation}
\frac{E_C^*}{E_C}=f(\alpha)e^{-\pi^2\alpha},
\label{general}
\end{equation} 
the pre-exponential factors differ widely. Using an instanton approach, Panyukov and Zaikin \cite{Panyukov_Zaikin} obtain $f(\alpha)\sim\alpha^2$, whereas Wang and Grabert \cite{Wang_Grabert} find $f(\alpha)\sim\alpha^3$. Lukyanov and Zamolodchikov \cite{Lukyanov} use a perturbative expansion in $1/\alpha$ and propose an expression similar to Ref.~\onlinecite{Wang_Grabert}, with an additional logarithmic term. The renormalization group (RG) calculation by Hofstetter and Zwerger \cite{Hofstetter} predicts a linear dependence, while the ``real-time RG'' approach of K\"onig and Schoeller \cite{Koenig} leads to a pre-factor $f(\alpha)\sim\alpha^{6.5}$ in the limit of large $\alpha$. We list these different theoretical predictions in Tab.~\ref{predictions}.

\begin{table}
\centering 
\caption{Analytical predictions for the leading correction to the exponential suppression, Eq.~(\ref{general}), of the effective charging energy in the limit of large $\alpha$.}
\begin{tabular}{lll}
\hline
\hline
Ref.\hspace{2mm} & Method & $f(\alpha)$ \\
\hline
\onlinecite{Panyukov_Zaikin} & instanton & $4\pi^4\alpha^2$ \\
\onlinecite{Wang_Grabert}    & instanton & $16\pi^6\alpha^3$ \\
\onlinecite{Lukyanov}        & perturbation theory\hspace{2mm} & $8\pi^6\alpha^3(1-(5\log\alpha)/(2\pi^2\alpha))$\\
\onlinecite{Hofstetter}      & RG & $\sim \alpha$\\
\onlinecite{Koenig}          & real-time RG & $0.5(\pi\alpha)^{6.5}$\\
\hline
\end{tabular}
\label{predictions}
\end{table}

Several attempts to resolve the issue by numerical simulations \cite{Hofstetter, Wang&Egger, Herrero} have failed because of size constraints and the inability to reach the asymptotic regime where the exponential factor in Eq.~(\ref{general}) dominates. Some of these Monte Carlo results even seemed to disagree for intermediate values of $\alpha$.

The purpose of this study is to provide accurate data far into the strong tunneling regime, enabling us to test the predictions from the various analytical approaches. We observe two distinct behaviors, depending on the inverse temperature $\beta$ and the tunneling strength $\alpha$. If $\beta E_C^* \ll 1$, multiple phase slip paths get suppressed and we find a pre-exponential factor which roughly agrees with the single-instanton predictions.\cite{Panyukov_Zaikin}  In the limit $\beta E_C^* \gg 1$, which is relevant for the comparison with analytical results for zero temperature, we find $f(\alpha)\sim \alpha^5$. This result disagrees with all the predictions in Tab.~\ref{predictions}. 

\section{Model and Method}

The partition function of the single electron box with gate charge zero can be written as a path integral over a compact angular variable $\phi$ (conjugate to the number of excess charges on the island)\cite{Schoen}
\begin{equation}
Z=\int_0^{2\pi} d\phi_0 \sum_{n=0}^\infty\int_{\phi_0}^{\phi_0\pm 2\pi n}\mathcal{D}\phi e^{-S[\phi]}.
\label{partitionfunction}
\end{equation}
In Eq.~(\ref{partitionfunction}), the sum is over all paths with winding number $\pm n$ and the imaginary time effective action reads ($\hbar=1$)
\begin{eqnarray}
S[\phi]&=&\frac{1}{4E_C}\int_0^\beta d\tau \left(\frac{d\phi}{d\tau}\right)^2\nonumber\\
&+&\alpha\int_0^\beta \int_0^\beta d\tau d\tau' \frac{(\frac{\pi}{\beta})^2 \sin^2(\frac{\phi(\tau)-\phi(\tau')}{2})}{\sin^2(\frac{\pi}{\beta}(\tau-\tau'))}.
\label{action}
\end{eqnarray} 
The first term accounts for the charging energy and the long ranged part for electron tunneling.

This system can be mapped to a chain of classical XY spins, as the discretized action becomes
\begin{equation}
S[\phi]=-\Gamma \sum_{j=1}^{N}\cos(\phi_{j+1}-\phi_j)-\alpha\sum_{j<j'}\frac{(\frac{\pi}{N})^2 \cos(\phi_j-\phi_{j'})}{\sin^2(\frac{\pi}{N}(j-j'))}.
\label{discreteaction}
\end{equation} 
We use a Trotter number $N$, corresponding to an imaginary time step (and short-distance cutoff) $\Delta\tau=\beta/N$ and nearest neighbor coupling $\Gamma=1/(2E_C\Delta\tau)$. Periodic boundary conditions $\phi_{N+1}=\phi_1$ are employed.

In Fourier space the action becomes local,
\begin{equation}
S[\phi]=\frac{1}{N}\sum_{k=0}^{N-1}g_k|\psi_k|^2,
\label{fourier_action}
\end{equation} 
where $\psi_k=\sum_{j=1}^{N}e^{i\frac{2\pi}{N}jk}e^{i\phi_j}$ denotes the Fourier transform of $e^{i\phi}$ and $g_k$ the Fourier transform of the kernel ($j\ne 0$)
\begin{equation}
g(j) = -\frac{\Gamma}{2}(\delta_{j,1}+\delta_{j,N-1})-\frac{\alpha}{2}\frac{(\pi/N)^2}{\sin((\pi/N)j)^2}.
\label{kernel}
\end{equation}
Using fast Fourier transformation \cite{FFTW} it is  possible to compute the action of a configuration in a time $O(N\log N)$ despite the long ranged interactions.

The winding number $\omega$ of a phase configuration $\{\phi_j\}_{j=1,\ldots N}$ is determined by summing up the phase differences as

\begin{equation}
\omega = \sum_{j=1}^{N}[(\phi_{j+1}-\phi_j+\pi)\mod 2\pi]-N\pi.
\label{phase_sum}
\end{equation}
The effective charging energy can then be computed from the expectation value of the winding number squared \cite{Hofstetter},

\begin{equation}
\frac{E_C^*}{E_C}=\frac{2\pi^2}{\beta E_C}\langle \omega^2\rangle.
\label{winding}
\end{equation}

\subsection{Cluster Monte Carlo}

In most of the following analysis, we use path integral Monte Carlo simulations to study the action in Eq.~(\ref{discreteaction}), employing a variant of the Swendsen-Wang cluster algorithm \cite{Swendsen&Wang, Wolff} to reduce auto-correlation times. The efficient treatment of long ranged interactions proposed by Luijten and Bl\"ote \cite{Luijten&Bloete} reduces the time for a cluster update to $O(N\log N)$. This allows us to simulate chains of up to $10^7$ spins, which is three orders of magnitude larger than the systems which have previously been studied.

We ran simulations for $\Delta\tau E_C = 0.5$, 0.125, 0.05 and 0.005 and found essentially the same results for the latter three values of the discretization step (especially for large $\alpha$, which is the limit of interest). We therefore consider an extrapolation in $\Delta\tau$ unnecessary for the purpose of this study and present data for $\Delta\tau E_C = 0.05$, unless otherwise stated. With this choice of $\Delta\tau$, the cluster algorithm allows us to simulate the single-electron box down to temperatures $\beta E_C\sim 10^4$. However, for large $\alpha$, the fraction of paths with winding number different from zero decreases as $\exp(-\pi^2\alpha)$. Due to this rapidly deteriorating efficiency, the cluster algorithm can only be used in the range $0\le \alpha \lesssim 3$.

\subsection{Transition Matrix Monte Carlo}

In order to compute the effective charging energy at even larger values of the tunneling conductance, we developed a new Monte Carlo approach, which attempts to insert or remove phase slips (kinks) and in doing so measures the relative weights of the different winding number sectors. As only paths of winding number 0 and 1 contribute in the limit of large $\alpha$, we restrict the discussion to the calculation of the relative weight of these two winding number sectors. However, the algorithm can also be used to study additional winding number sectors.

We denote by $w_n$ the weight of the configurations with winding number $\pm n$,

\begin{equation}
w_n=\frac{\int_0^{2\pi}d\phi_0\int_{\phi_0}^{\phi_0\pm2\pi n}\mathcal{D}\phi e^{-S[\phi]}}{\int_0^{2\pi}d\phi_0\sum_{n=0}^{\infty}\int_{\phi_0}^{\phi_0\pm2\pi n}\mathcal{D}\phi e^{-S[\phi]}}
\label{weight}
\end{equation}
The effective charging energy (\ref{winding}) can then be expressed for large $\alpha$ as

\begin{equation}
\frac{E_C^*}{E_C} = \frac{2\pi^2}{\beta E_C}\frac{\sum_{n=0}^{\infty} n^2 w_n}{\sum_{n=0}^\infty w_n}\approx\frac{2\pi^2}{\beta E_C}\frac{w_1}{w_0}.
\label{large_winding}
\end{equation}

The weights $w_0$ and $w_1$ correspond to a ``density of states" in winding number space. In order to determine $w_1/w_0$ we could sample the winding number space with the inverse density of states and energy space with the usual Boltzmann weight. The acceptance probability for a proposed kink update from a configuration with winding number $n$ and action $S$ to winding number $n'$ and action $S'$ then becomes

\begin{equation}
P((n,S)\rightarrow(n',S'))=\min\left(1,\frac{w_n}{w_{n'}}\exp(S-S')\right),
\label{probability_metro}
\end{equation}
whereas a cluster update from $n$ to $n'$ would be accepted with probability

\begin{equation}
P((n)\rightarrow(n'))=\min\left(1,\frac{w_n}{w_{n'}}\right).
\label{probability_sw}
\end{equation}
The relative density of states $w_1/w_0$ then satisfies the ``flat histogram" condition, which for $n\in\{0,1\}$ can be expressed as

\begin{equation}
\left\langle P(0\rightarrow 1)\right\rangle(w_1/w_0) = \left\langle P(1\rightarrow 0)\right\rangle(w_1/w_0).
\label{flat_histogram}
\end{equation}
Averages are taken over a sequence of kink- and cluster updates.

Our strategy is to determine the transition probabilities $\left\langle P(0\rightarrow 1)\right\rangle$ and $\left\langle P(1\rightarrow 0)\right\rangle$ rather than trying to find the relative occupation of the winding sectors $w_1/w_0$ by a Wang-Landau type iterative procedure \cite{Wang&Landau}. This is an approach in the spirit of the transition Matrix Monte Carlo method \cite{Wang&Swendsen}.

Each winding number sector is treated separately and the cluster updates serve essentially to randomize the configurations within that sector, although their contribution to the probabilities $\langle P(i\rightarrow j)\rangle$ becomes important for smaller $\alpha$. In a kink update we insert a phase slip

\begin{equation}
  \phi_{\text{kink}}^{(\tau_0, \lambda)} = \pm 2\arctan\left(\frac{\tau-\tau_0}{\lambda}\right)
\label{kink}
\end{equation}
of random width $\lambda$ at some random position $\tau_0$ (or rather $\phi_{\text{kink}}^{(\tau_0, \lambda)} = \pm \pi\arctan((\tau-\tau_0)/\lambda)/\arctan(\beta/2\lambda)$, periodically continued outside the the interval $-\frac{\beta}{2}\le \tau-\tau_0\le\frac{\beta}{2}$, which is a slightly modified version compatible with the finite size of the system). These are stationary paths of the long ranged part of the action (\ref{action}) in the limit $\beta\rightarrow\infty$.

\begin{figure}[t]
\centering
\includegraphics [angle=-90, width= 9cm] {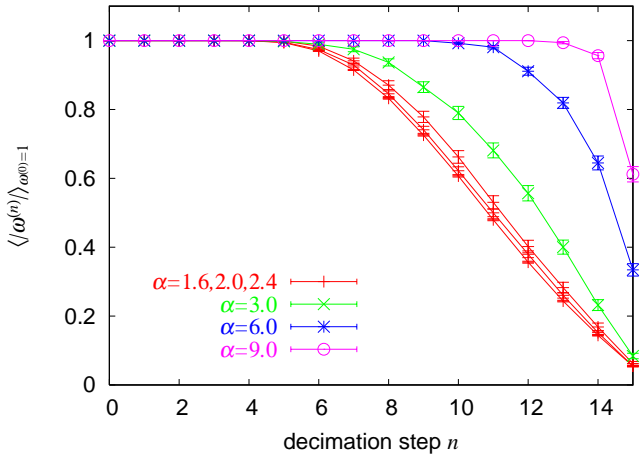}
\caption{Average winding number $\langle |\omega^{(n)}| \rangle$ of coarse grained configurations. The original configuration has winding number $|w^{(0)}|=1$. It is renormalized by removing every second spin. Kinks of width $\approx 2^n$ will disappear after approximately $n$ iterations, resulting in a coarse-grained configuration of winding number 0.}
\label{decimation}
\end{figure} 

It is important to consider the entire range of widths including stretched out configurations. In order to study the typical distribution of widths we generated configurations for $N=10^5$ ($\beta E_C=5\cdot 10^3$) in winding sector 1 using the cluster algorithm and applied a decimation procedure in which the number of spins was reduced by half at each step. The average winding number $\langle |\omega^{(n)}|\rangle$ of these coarse-grained configurations is shown as a function of decimation steps $n$ in Fig.~\ref{decimation}. Removing every second spin reduces the width of the kink to approximately half its previous value until the kink disappears completely. From the dependence of $\langle |\omega^{(n)}|\rangle$ on $n$ it is therefore possible to gain insight into the original distribution of kink widths in winding sector 1. One finds from Fig.~\ref{decimation}, where we show the average winding after coarse-graining a configuration, that the kink widths for $1.6\le \alpha \le 2.4$ are approximately uniformly distributed in the range $[2^6, 2^{16}]$, whereas for $\alpha=9$ only smooth configurations with widths $\in[2^{13}, 2^{16}]$ are generated. The sharpest kinks have a width of $2^5$. In order to cover the whole range of relevant paths, we chose $\lambda_\text{min} = 10\beta/N$ and $\lambda_\text{max}=\beta$. 

Inserting $\phi_{\text{kink}}^{(\tau_0, \lambda)}$ changes the original configuration $\phi$ with winding number $n$ to $\phi'$ with winding number $n\pm 1$. The corresponding Boltzmann weights $\exp(S[\phi]-S[\phi'])$ are computed using Eq.~(\ref{fourier_action}) and stored during the simulation. Updates which produce a configuration with winding number $|n|>1$ as well as cluster updates which leave the winding number unchanged get the weight 0. A cluster update which connects to the other winding number sector gets the weight 1.  From these data, one can calculate the average transition probabilities $\langle P(i\rightarrow j)\rangle$ as a function of $w_1/w_0$ using Eqs.~(\ref{probability_metro}) and (\ref{probability_sw}). Finally, the relative weight of winding sector 1 is determined by solving Eq.~(\ref{flat_histogram}).
   
\begin{figure}[t]
\centering
\includegraphics [angle=-90, width= 9cm] {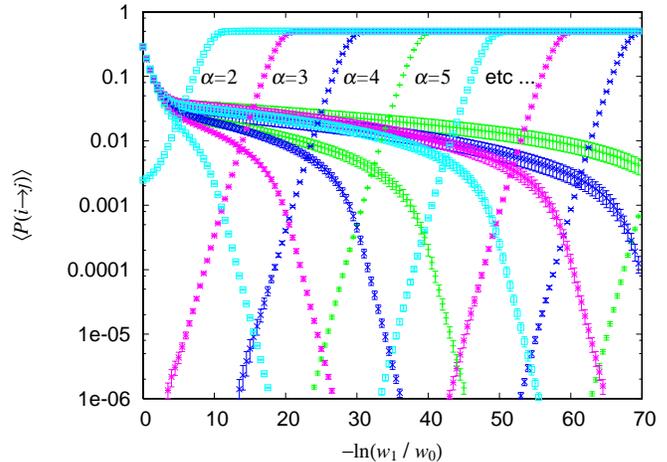}
\caption{Transition probabilities $\langle P(0\rightarrow 1)\rangle$ (with positive slope) and $\langle P(1\rightarrow 0)\rangle$ (with negative slope) plotted as a function of $-\ln(w_1/w_0)$. The intersection points of these curves determines $w_1/w_0$. From left to right, the data correspond to $\alpha=2$, 3, $\ldots$, 9.}
\label{intersect}
\end{figure} 

\begin{figure}[t]
\centering
\includegraphics [angle=-90, width= 9cm] {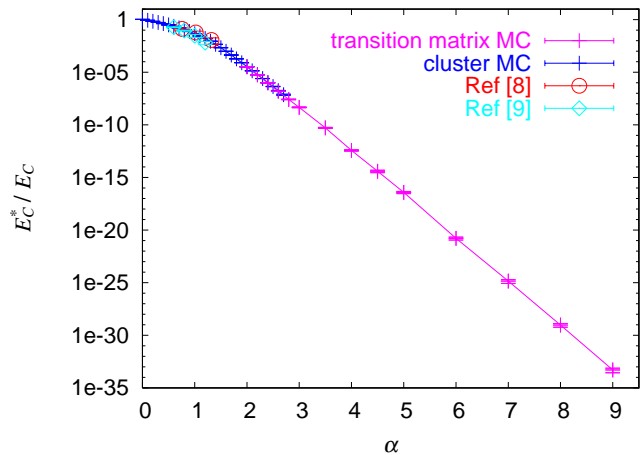}
\caption{Effective charging energy for $\beta E_C=5\cdot 10^3$ computed using cluster updates and kink updates. Previously published results obtained by means of local update schemes are shown for comparison.}
\label{local_cluster_kink}
\end{figure} 

To illustrate this procedure, Fig.~\ref{intersect} shows the probabilities $\langle P(0\rightarrow 1)\rangle$ (with positive slope) and $\langle P(1\rightarrow 0)\rangle$ (with negative slope) as a function of $-\ln(w_1/w_0)$ for several values of $\alpha$. The intersection points of these curves determine $w_1/w_0$. The effective charging energies obtained by this new method are consistent with the results from cluster Monte Carlo for the values of $\alpha$ which can be treated with the latter method. The new approach, however, allows us to simulate the system at much higher tunneling conductance, as it improves the sensitivity to winding number fluctuations by dozens of orders of magnitude.

\section {Effective Charging Energy}

\subsection{Leading exponential behavior}

The range of applicability of the different simulation methods is illustrated in Fig.~\ref{local_cluster_kink}, where we plot the effective charging energy as a function of tunneling conductance for $\beta E_C = 5\cdot 10^3$. While cluster updates allow to compute accurate data at small and intermediate values of $\alpha$, they lose their effectiveness around $E_C^*/E_C\approx 10^{-8}$ or $\alpha\approx 3$. On the other hand, the kink-updates enable us to cover tunneling strengths beyond $\alpha = 9$, which corresponds to $E_C^*/E_C\approx 10^{-34}$. This highly sensitive method allows us to clearly observe the theoretically predicted exponential suppression (\ref{general}) of the effective charging energy. For comparison, we also plot some results from local update simulations.\cite{Wang&Egger, Herrero} Although these simulations were performed on much smaller lattices, they merely reached $E_C^*/E_C \approx 10^{-2}$.

\subsection{Weak and intermediate tunneling}

\begin{figure}[t]
\centering
\includegraphics [angle=-90, width= 9cm] {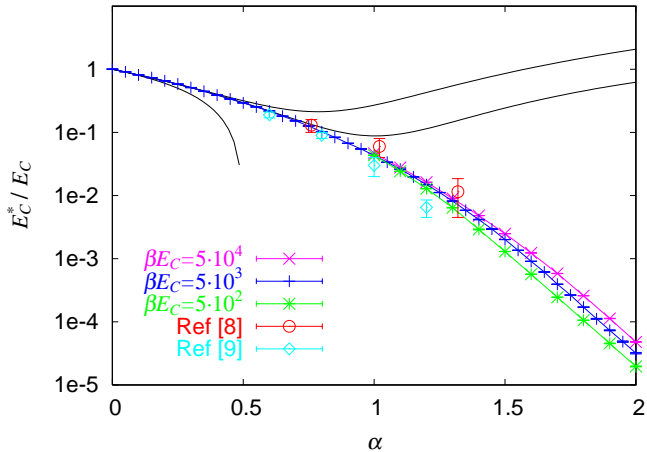}
\caption{Effective charging energy in the weak and intermediate tunneling regime. The black lines show the first-, second- and third-order perturbation results (\ref{perturbation_series}). The data for $\beta E_C=5\cdot 10^3$ were computed with $\Delta \tau=0.005$ in order to reduce the energy shift due to discretization effects.}
\label{perturbation}
\end{figure}

While the weak tunneling regime can be treated perturbatively, few theoretical results exist for the experimentally accessible intermediate region. Since previous Monte Carlo simulations\cite{Herrero, Wang&Egger, Hofstetter} seemed to disagree in this region, we plot our data for weak and intermediate tunneling in Fig.~\ref{perturbation}. The third order perturbation result (\ref{perturbation_series}) reproduces our data for $\beta E_C = 5\cdot 10^3$ and $\Delta\tau=0.005$ up to $\alpha=0.6$ (the data for $\Delta\tau=0.05$ are shifted to slightly larger $E_C^*/E_C$). Also shown in Fig.~\ref{perturbation} are results for lower and higher temperature and the data of Wang \textit{et al.} \cite{Wang&Egger} and Herrero \textit{et al.}\cite{Herrero}. The results in Ref.~\onlinecite{Wang&Egger} were obtained for $\beta E_C=5\cdot 10^2$ and are consistent with our data. The larger time step $\Delta\tau E_C=0.2$ used in their simulation may be the reason for the tendency to somewhat larger values of $E_C^*$. The systematically lower values in Ref.~\onlinecite{Herrero} were computed at higher temperature $\beta E_C = 10^2$, which explains the discrepancy.

Around $\alpha\approx 0.6$ the system enters an intermediate tunneling regime, where perturbation theory fails but the asymptotic formulas for large tunneling are not yet valid. The exponential suppression of the effective charging emerges above $\alpha\approx 1.5$ (see Figs.~\ref{local_cluster_kink} and \ref{perturbation}). This strong tunneling region cannot be accessed by local update schemes.

\subsection{Strong tunneling and crossover}

In order to determine the pre-exponential factor $f(\alpha)$ in Eq.~(\ref{general}), we multiply the data with $\exp(\pi^2\alpha)$. The value $\pi^2\alpha$ corresponds to the (long ranged) action of an optimal phase slip path (\ref{kink}) in the limit $\beta\rightarrow \infty$ and $\Delta\tau\rightarrow 0$. Since the leading power of the pre-exponential factor will depend sensitively on this value, one might wonder whether the discretization leads to a significant change. For our values of $\beta$ and $\Delta\tau$, however, we found that periodic versions of the paths (\ref{kink}), with  $\lambda$ not too small, have an action which deviates very little from $\pi^2\alpha$.

\begin{figure}[t]
\centering
\includegraphics [angle=-90, width= 9cm] {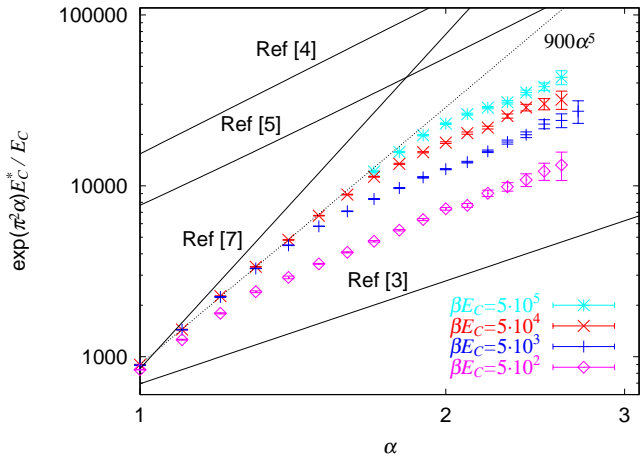}
\caption{Correction to the exponential behavior, Eq.~\ref{general}, at intermediate and large tunneling conductance and comparison with analytical predictions. A crossover from intermediate tunneling (zero temperature) behavior to strong tunneling (intermediate temperature) behavior occurs at $\beta E_C^*\approx 1$. }
\label{slope}
\end{figure}

In Fig.~\ref{slope}, we plot cluster Monte Carlo results for $\beta E_C = 5\cdot 10^2$, $5\cdot 10^3$, $5\cdot 10^4$ and $5\cdot 10^5$. The lowest temperature was computed with $\Delta\tau E_C=0.2$, the others with $\Delta\tau E_C=0.05$. Also shown in Fig.~\ref{slope} are the asymptotic predictions from several analytical calculations. The line with slope 2 (in the log-log plot) is the result of Panyukov and Zaikin\cite{Panyukov_Zaikin}, those with slope 3 were predicted in Refs.~\onlinecite{Wang_Grabert} and \onlinecite{Lukyanov} while the line with slope 6.5 shows the asymptotic form of the``real time RG" result by K\"onig and Schoeller\cite{Koenig}. Hofstetter and Zwerger\cite{Hofstetter} predict a slope of 1, but not the amplitude.

As the tunneling strength increases, the effective charging energy $E_C^*$ gets reduced and for fixed temperature one observes a crossover from intermediate tunneling behavior, where $\beta E_C^*\gg 1$, to strong tunneling behavior in the region $\beta E_C^*\ll 1$, or equivalently for $\alpha$ fixed, from zero temperature behavior to an intermediate temperature behavior. This crossover occurs around $\beta E_C^*=1$ and is clearly visible in Fig.~\ref{slope}. For the temperatures plotted in this figure, the crossover condition corresponds to $\alpha$-values in the range $1.4\le \alpha \le 2.2$.

\subsection{Zero temperature regime, $\beta E_C^* \gg 1$}

Theoretical predictions for zero temperature have to be compared to the Monte Carlo data in the zero temperature (intermediate tunneling) limit $\beta E_C \gg \beta E_C^*\gg 1$. Because of size restrictions we can only obtain accurate results for $\alpha\lesssim 1.6$. These data points can be fitted with a power-law as indicated by the dotted line in Fig.~\ref{slope}. It corresponds to a pre-exponential factor
\begin{equation}
f(\alpha)\sim \alpha^{5}.
\label{weak_factor}
\end{equation}

\begin{figure}[t]
\centering
\includegraphics [angle=-90, width= 9cm] {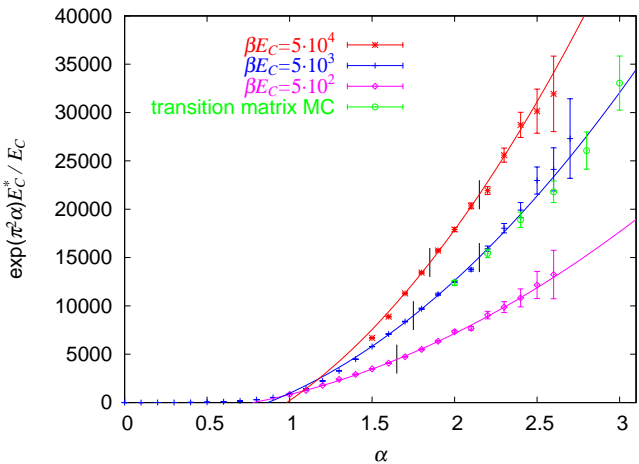}
\caption{The pre-exponential factors in the strong tunneling regime can be fitted by shifted parabolas, Eq.~(\ref{strong_factor}). The vertical lines indicate the values of $\alpha$, where the occupation of winding number sectors 1 and 2 vanish (see Fig.~(\ref{sectors})). Data points marked with circles show the results for $\beta E_C=5\cdot 10^3$ computed by the transition matrix method, which considers configurations with kink-anti-kink pairs.}
\label{parabolic}
\end{figure}
\begin{figure}[t]
\centering
\includegraphics [angle=-90, width= 9cm] {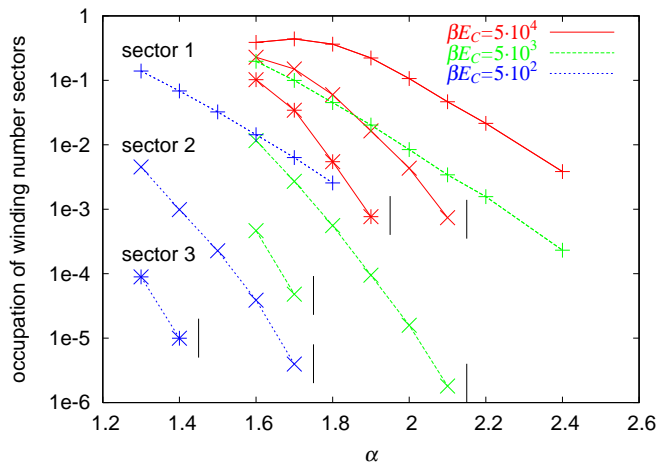}
\caption{Relative occupation of the winding number sectors 1, 2 and 3 as a function of tunneling strength. As $\alpha$ increases, higher-winding-number configurations get suppressed and in a finite number of updates, multiple kink paths will not be generated at large values of $\alpha$.}
\label{sectors}
\end{figure}

Because this fit is performed in the region of intermediate tunneling strengths, it is possible that the slope of the zero-temperature curve will decrease and eventually approach for example the solution of Lukyanov and Zalomodchikov.\cite{Lukyanov} The calculation of Ref. \onlinecite{Koenig}  overestimates the exponent, whereas the result of Ref. \onlinecite{Panyukov_Zaikin} lies too low. It is even more unlikely that the zero-temperature curve will approach the slope 1 predicted in Ref. \onlinecite{Hofstetter}. The latter RG-calculation was based on the assumption that the winding number remains unchanged under renormalization, which appears incorrect, as can be seen in Fig.~\ref{decimation}.

\subsection{Intermediate temperature regime, $\beta E_C^* \ll 1$}

In the intermediate temperature (strong-tunneling) region $\beta E_C \gg 1 \gg \beta E_C^*$, the pre-exponential factor can be fitted by a shifted parabola
\begin{equation}
f(\alpha)\sim \alpha^2+\text{const}.
\label{strong_factor}
\end{equation}

In Fig.~\ref{parabolic} we show these parabolic fits for $\beta E_C=5\cdot 10^2$, $5\cdot 10^3$ and $5\cdot 10^4$. Upon close inspection, these data show some intriguing features. At certain values of $\alpha$, the slope seems to change suddenly and the quality of the data deteriorates. This feature is a result of the suppression of multiple-kink paths. As $\alpha$ is increased, the occupation of higher winding sectors is reduced and for a finite number of measurements it will eventually vanish. The visible change in slope occurs at the value of $\alpha$, where multiple-kink paths get completely suppressed.

\begin{figure}[t]
\centering
\includegraphics [angle=-90, width= 9cm] {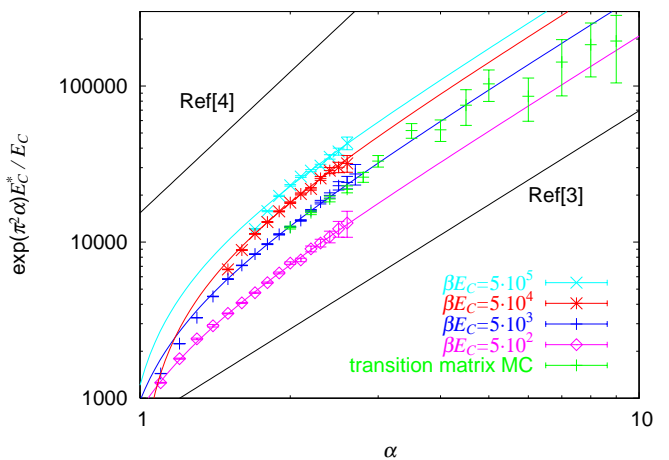}
\caption{Pre-exponential factor of the effective charging energy in the intermediate and strong tunneling regime. The parabolic fit to the cluster Monte Carlo data (see Fig.~\ref{parabolic}) is consistent with the data obtained from the transition matrix approach even at much higher tunneling conductance.}
\label{large_alpha}
\end{figure}

To illustrate this point, we plot in Fig.~\ref{sectors} the relative occupation of the lowest three winding number sectors as a function of $\alpha$. The position where the winding numbers 2 and 3 get suppressed are marked by the vertical line segments in Figs.~\ref{parabolic} and \ref{sectors}. The transition matrix Monte Carlo data for very large tunneling strength are consistent with the parabolic fit to the cluster Monte Carlo data, as shown in Fig.~\ref{large_alpha}.

The instanton calculations \cite{Panyukov_Zaikin, Wang_Grabert} neglect the interactions between phase-slips. This is probably the reason why they predict a pre-exponential factor which is roughly consistent with the simulation results in the large tunneling (intermediat temperature) region, where multiple phase slips no longer contribute. In the zero temperature regime $\beta E_C^*\gg 1$, however, multiple-kink configurations are abundant (see Fig.~\ref{sectors}). The considerably different slope in this region leads to the conclusion, that interactions between phase slips are important. They should be considered in a theory which attempts to predict the pre-exponential factor $f(\alpha)$ for the single-electron box at zero temperature. 

\section{Conclusions}

We calculated the effective charging energy of a single-electron box using cluster Monte Carlo simulations and a new transition matrix Monte Carlo approach. The new method allows us to study very large tunneling strengths and to observe the leading exponential suppression of the charging energy, $E_C^*/E_C=f(\alpha)\exp(-\pi^2\alpha)$, over more than thirty orders of magnitude. For a finite inverse temperature $\beta$ one observes two regimes. In the large tunneling (intermediate temperature) regime, $E_C^*\beta \ll 1$, where multiple kink paths are suppressed, the pre-exponential factor has a leading power $f(\alpha)\sim \alpha^2$, consistent with the instanton calculation of Ref. \onlinecite{Panyukov_Zaikin}. In the intermediate tunneling (zero temperature) regime, $E_C^*\beta\gg 1$, multiple kink paths are abundant and the interactions between phase-slips become important. Simulation results obtained in this regime can be compared to analytical predictions for zero temperature. Fitting our data in the range $1.0<\alpha<1.6$ to a power-law, we find a pre-exponential factor $f(\alpha)\sim \alpha^{5}$, which is not consistent with any of the theories.

\acknowledgements

We acknowledge support by the Swiss National Science Foundation, the Aspen Center for Physics, and the KITP in Santa Barbara. The calculations have been performed on the Asgard Beowulf cluster at ETH Z\"urich, using the ALPS library \cite{ALPS}. We thank P.~de~Forcrand,  Ch.~Helm,  W. Hofstetter, R.~Sch\"afer and W. Zwerger for helpful discussions.

\end{document}